\documentclass[english,twocolumn,showpacs,aps,prd]{revtex4}
\pdfoutput=1

\usepackage{preprintcover}  
\PreprintCoverPaperTitle{Search for microscopic black holes in a like-sign dimuon final state using large track multiplicity with the ATLAS detector}  
\PreprintIdNumber{CERN-PH-EP-2013-120}  
\PreprintCoverAbstract{
A search is presented for microscopic black holes in a like-sign dimuon final state in proton--proton collisions at $\sqrt{s} = 8\tev$. The data were collected with the ATLAS detector at the Large Hadron Collider in 2012 and correspond to an integrated luminosity of 20.3~\ifb. Using  a high track multiplicity requirement, $0.6 \pm 0.2$ background events from Standard Model processes are predicted and none observed.  This result is interpreted in the context of low-scale gravity models and 95\% CL lower limits on microscopic black hole masses are set for different model assumptions.
}  
\PreprintJournalName{Phys. Rev. D}  

\usepackage{graphicx} 
\graphicspath{{figures/}}
\DeclareGraphicsExtensions{.pdf}
\usepackage{atlasphysics} 
\usepackage{rotating} 
\usepackage{multirow}
\usepackage{subfigure}
\usepackage{hyperref}

\begin{document}

\vspace{0.5cm}

\title{Search for microscopic black holes in a like-sign dimuon final state using large track multiplicity with the ATLAS detector}

\author{The ATLAS Collaboration}

\date{\today}

\begin{abstract}
A search is presented for microscopic black holes in a like-sign dimuon final state in proton--proton collisions at $\sqrt{s} = 8\tev$. The data were collected with the ATLAS detector at the Large Hadron Collider in 2012 and correspond to an integrated luminosity of 20.3~\ifb. Using  a high track multiplicity requirement, $0.6 \pm 0.2$ background events from Standard Model processes are predicted and none observed.  This result is interpreted in the context of low-scale gravity models and 95\% CL lower limits on microscopic black hole masses are set for different model assumptions.

\end{abstract}


\maketitle

\section{Introduction}
\label{sec:intro}

The hierarchy problem, in which the Planck scale ($\MP \approx 10^{19}\GeV$) is much higher than
the electroweak scale ($\approx 100$~\GeV), provides a strong motivation to search for new phenomena
not described by the Standard Model of particle physics. A theory introducing extra dimensions is one possible solution. 
In some models of extra dimensions,
the gravitational field propagates into $n+4$ dimensions, where $n$ is the number of extra dimensions
beyond the four space-time dimensions. 
One of the models of extra dimensions is the model proposed by Arkani-Hamed, Dimopoulos, 
and Dvali (ADD)~\cite{ArkaniHamed:1998rs,Antoniadis:1998ig,ArkaniHamed:1998nn}, in which the gravitational field propagates into large, flat,
extra dimensions while the Standard Model particles are localized in four space-time dimensions.
Since the gravitational field propagates into the extra dimensions, it is measured at a reduced strength in the four space-time dimensions. 
Thus, the fundamental  Planck scale in 
$D=4 + n$ dimensions, \MD, could be comparable with the electroweak scale.

If extra dimensions exist and \MD{} is of the order of 1~\TeV{}, microscopic black holes with \TeV-scale mass could exist and 
be produced at the Large Hadron Collider (LHC)~\cite{Argyres:1998qn,Banks:1999gd,Dimopoulos:2001hw,Giddings:2001bu,Kanti:2004nr}. 
These black holes are produced when the impact parameter of the two colliding protons is smaller than the higher-dimensional
event horizon of a black hole with mass equal to the invariant mass of the colliding proton system.
 
The black hole production has a continuous mass distribution ranging from \MD\ to the proton--proton center-of-mass energy.  The black holes evaporate by emitting Hawking radiation~\cite{Hawking:1974sw}, which determines the energy and
multiplicity of the emitted particles. 
The relative multiplicities of different particle types are determined by the number of degrees of freedom of each particle type
and the decay modes of the emitted unstable particles. Black hole events are thus expected to have a high multiplicity of high-momentum particles.

This paper describes a search for black holes in a like-sign dimuon final state. This final state can arise from muons directly produced by the black hole, or from the decay of
Standard Model particles produced by the black hole. The final state is expected to have low Standard Model
backgrounds while retaining a high signal acceptance. Since the microscopic black holes can decay to a large number of particles with high transverse momentum (\pt), 
the total track multiplicity of the event is exploited to distinguish signal events from backgrounds. The final result is obtained from 
the event yield
in a signal region defined by high track multiplicity.

The following assumptions and conventions apply in this analysis.
The classical approximations used for black hole production, and the semi-classical approximations for the decay are predicted to be valid only for
black hole masses well above \MD. A lower threshold (\MTH) is applied to the black hole mass, $\MTH > \MD + 0.5 \TeV$, to reduce contributions from regions where the models
are invalid, and the production cross section is set to zero if the proton--proton center of mass energy is below \MTH.
The mass of the produced black hole decreases from \MTH\ to \MD\ as a result of the emission of Hawking radiation.
When the mass of the black hole approaches $\MD$,
quantum gravity effects become important. In the final stage of the black hole decay, 
classical evaporation is no longer a good description. In such cases where the black hole mass is near \MD, 
the burst model adopted by the \blackmax\ event generator~\cite{Dai:2007ki} is used in the final part of the decay. 
No graviton initial-state radiation or emission from the black hole is considered in this paper.
Models of rotating and non-rotating black holes are both studied. The track multiplicity is predicted to be slightly lower for rotating
black holes~\cite{bhrot}.

A previous result from ATLAS~\cite{dimu2011} in this final state excludes at 95\% confidence level (CL) the production of black holes with 
$\MTH \le$ 3.3, 3.6 and 3.7~\TeV{} for \MD{} of 1.5~\tev{} and for $n =$ 2, 4, and 6, respectively.
A previous search by the ATLAS Collaboration in a lepton+jets final state~\cite{ljets2011} excludes at 95\% CL black holes with $\MTH \le 4.5~\TeV$ 
for $\MD=1.5$~\tev{} and $n = 6$. 
The CMS Collaboration has conducted searches in a multi-object final state and excluded the production of black holes at 95\% CL
with $\MTH \le$  5.7, 6.1, and 6.2~\TeV{} for $\MD= 1.5$~\TeV\ and for $n =$ 2, 4, 
and 6, respectively~\cite{cmsbh,cmsbh7tev}. 

The rest of this paper is organized as follows. A brief description of the ATLAS detector is given in Sec.~\ref{sec:atlas}. The data
and simulation samples are described in Sec.~\ref{sec:samples}, followed by the event selection in Sec.~\ref{sec:eventselection}.
The background estimation techniques are discussed in Sec.~\ref{sec:backgrounds}. The final results and their interpretation
are presented in Sec.~\ref{sec:results}.

\section{The ATLAS Detector}
\label{sec:atlas}

The ATLAS detector~\cite{atlas} is a multi-purpose detector with a forward-backward 
symmetric cylindrical geometry covering nearly the entire solid angle~\footnote{ATLAS uses a right-handed coordinate system with its origin at the nominal 
interaction point (IP) in the center of the detector and the $z$-axis along the beam pipe. 
The $x$-axis points from the IP to the center of the LHC ring, and the $y$-axis points upward. Cylindrical coordinates $(r,\phi)$ are used in the transverse plane, 
$\phi$ being the azimuthal angle around the beam pipe. The pseudorapidity is defined in terms of the polar angle $\theta$ as $\eta=-\ln\tan(\theta/2)$.} 
around the collision point with layers of tracking detectors, calorimeters and muon chambers. The inner detector is immersed in 
a 2~T axial magnetic field, provided by a solenoid, in the $z$ direction and provides charged particle tracking in the 
pseudorapidity range $|\eta|<2.5$. A silicon pixel
detector covers the luminous region and typically provides three measurements per track, followed by a silicon microstrip
tracker (SCT) that provides measurements from eight strip layers. In the region with $|\eta|<2.0$, the silicon detectors are complemented by a
transition radiation tracker (TRT), which provides more than 30 straw-tube measurements per track.

The calorimeter system covers the range $|\eta| < 4.9$. Lead/liquid argon (lead/LAr) electromagnetic sampling calorimeters
cover the range $|\eta| < 3.2$, with an additional thin lead/LAr presampler covering $|\eta| < 1.8$ to correct for energy loss in
material upstream of the calorimeters. Hadronic calorimetry is provided by a steel/scintillator-tile calorimeter 
over $|\eta| < 1.7$  and two copper/LAr endcap calorimeters over $1.75< |\eta| < 3.2$. The solid
angle coverage is completed with forward copper/LAr and tungsten/LAr calorimeters for electromagnetic and hadronic measurements 
respectively up to $|\eta|$ of 4.9.

The muon spectrometer consists of separate trigger and high-precision tracking chambers that measure the deflection of muon tracks
in a magnetic field with a bending integral in the range of 2~T\,m to 8~T\,m. The magnetic field is generated by three
superconducting air-core toroid magnet systems. The tracking chambers cover the region $|\eta|<2.7$ with three layers of monitored drift tubes
 supplemented by cathode strip chambers in the innermost region of the endcap muon spectrometer.
The muon trigger system covers the range $|\eta|<2.4$ with resistive plate chambers in the barrel, and thin gap chambers in the
endcap regions.

\section{Data and Monte Carlo Samples}
\label{sec:samples}
The data used in this analysis were collected with the ATLAS detector from proton--proton collisions produced at $\sqrt{s} = 8 \tev$ in 2012.
The data correspond to an integrated luminosity of 20.3~\ifb.
The uncertainty on the luminosity is 2.8\% and is derived, following the same methodology as that detailed in 
Ref.~\cite{lumi2011}, from a preliminary calibration of the luminosity scale obtained from beam-separation scans 
performed in November 2012.
The events used for this analysis were recorded with a single-muon trigger with a threshold at $36~\gev$ on the muon \pt. 
The single muon trigger efficiency reaches a plateau for muons with $\pt > 40 \gev$ and 
the plateau efficiency is $71\%$ in the barrel and $87\%$ in the endcap for muons reconstructed offline.
The inefficiency in the trigger is driven mainly by the uninstrumented regions of the muon trigger system.

Monte Carlo (MC) samples are used for both signal and background modeling.
The ATLAS detector is simulated using \geant~\cite{geant}, and simulation samples~\cite{atlassim} are
reconstructed using the same software as that used for the collision data. The effect of additional proton--proton collisions in the same or neighboring
bunch crossings is modeled by overlaying simulated minimum-bias events onto the original hard-scattering event. 
MC events are then re-weighted so that the reconstructed vertex multiplicity distribution agrees with the one from data.

The dominant background processes are top-quark pair (\ttbar), diboson, and $W$+jets production with smaller contributions from single-top production. 
Background MC samples for the $\ttbar$ and the single-top ($Wt$-channel) processes are generated
using \powheg~\cite{powheg} and the CT10~\cite{ct10} parton distribution functions (PDFs).
Fragmentation and hadronization of the events is done with \pythia\ v6.426~\cite{pythia} using the Perugia tune~\cite{perugia}.
The top-quark mass is fixed at 172.5$\gev$. Alternative samples for studying the systematic uncertainty are made using the 
\alpgen\ v2.14~\cite{alpgen} or
\mcatnlo\ v4.03~\cite{mcatnlo}  generators with \herwig\ v6.520~\cite{herwig} used for hadronization and \jimmy\ v4.31~\cite{jimmy} used to model
the underlying event for both generators. The nominal single-top
sample uses the diagram-removal scheme~\cite{Frixione:2008yi} and an alternative sample using the diagram-subtraction scheme 
is produced for systematic studies. 
Diboson samples ($WZ$ and $ZZ$) are generated and hadronized using \sherpa\ v1.4.1~\cite{sherpa}.
The diboson samples are produced with the CT10 PDF set and use the ATLAS Underlying Event Tune 2B (AUET2B)~\cite{auet2b}. These samples include the case where the $Z$ boson (or $\gamma^*)$ is off-shell, with the invariant mass of the $\gamma^*$ required to be above twice the muon mass.

Signal MC samples are generated using \blackmax\ v2.02~\cite{Dai:2007ki,Dai:2009by} and the MSTW 2008 LO~\cite{Martin:2009iq} PDF set
with the mass of the black hole used as the factorization and renormalization scale. The signal samples are hadronized with 
\pythia\ v8.165~\cite{pythia8} using the AUET2B tune. Signal samples for rotating and non-rotating black holes are produced
by varying \MD\ between 1.0~\TeV\ and 4.5~\TeV, and \MTH\ between 3.0~\TeV\ and 6.5~\TeV. In each case, samples are generated
with $n$ = 2, 4, and 6. As an illustration, the expected yield from rotating black holes in a model with
$n$ = 4, $\MTH = 5 \TeV$, and $\MD = 1.5 \TeV$ is shown throughout the paper.

\section{Event Selection}
\label{sec:eventselection}
Events in data passing the single-muon trigger are selected for this analysis. The
detector was required to have been operating properly when these events were
collected. 
Events are also required to have a primary vertex reconstructed from at
least five tracks with $\pt > 400 \mev$. In events with multiple vertices, 
the vertex whose associated tracks have the largest $\Sigma{p_{\mathrm{T}}^{2}}$ is identified as the primary vertex.

Muon candidates are
reconstructed from tracks measured in the muon spectrometer (MS). The MS tracks
are matched with inner detector (ID) tracks using a procedure that takes material
effects into account. The final parameters for the muon candidates are obtained
from a statistical combination of the measured quantities in the MS and the ID.
The muon candidates must satisfy $|\eta|<2.4$ and have $\pt > 15 \gev$.
The quality of the ID track associated with a muon is ensured by imposing requirements~\cite{muonperf}
on the number of pixel, SCT and TRT hits associated with the track.
The ID tracks must also pass a requirement on 
the longitudinal impact parameter ($z_0$) with respect to the primary vertex, $|z_0\sin\theta| < 1.5\,{\rm mm}$.

Events are required to have at least two muons. The  two muons with the highest  \pt\ are
required to have the same charge. The muon with the highest \pt\ in the event is called the leading muon, while the muon with the
second highest \pt\ is called the subleading muon.
The leading muon is required to satisfy $\pt > 40 \gev$ to be above the trigger threshold and
pass requirements on isolation and transverse impact parameter as described below. 
No such requirements are made for the subleading muon.

The muon isolation is constructed from the sum of transverse momenta of other ID tracks in a cone 
in $\eta$--$\phi$ space of radius ${\Delta R} = \sqrt{(\Delta\eta)^2 + (\Delta\phi)^2} = 0.2$ around the muon. For the leading muon, the sum
is required to be less than 20\% of the muon \pt. The impact parameter significance for the muons is
defined as $|d_{0}/\sigma(d_{0})|$, where $d_0$ is the transverse impact parameter of the muon, and $\sigma(d_0)$ is the associated uncertainty.
The leading muon must satisfy $|d_{0}/\sigma(d_{0})| \le 3.0$.
The leading and subleading muons are required to be separated by ${\Delta R} > 0.2$.

The total track multiplicity (\ntrk) of the event is calculated by considering all ID tracks with
$\pt > 10 \gev$ and $|\eta|<2.5$ that pass the same quality and $z_0$ criteria as those for the muon ID tracks.
The track selection is thus less stringent than the muon selection and the track multiplicity counts
the two muons as well.

All selections except the trigger requirements are applied to the MC events. 
The MC events are assigned a weight based on their probability to pass the trigger requirements. The total probability is calculated by considering
each muon in the event and the individual probability of the muon to pass the trigger selection.
The MC events are also corrected to account for minor differences between data and MC simulation in the muon reconstruction and identification
efficiencies by applying $\pt$- and $\eta$-dependent scale factors. The tracking efficiency in MC simulation~\cite{minbias} is consistent with data
and has been confirmed with additional studies of tracking performance in a dense environment~\cite{Aad:2011sc}.
Thus no corrections are applied to the simulation for tracking performance.

Signal and validation regions are defined after the like-sign dimuon preselection described above and using the \ntrk\ definition.
The signal region is defined as follows:
\begin{itemize}
\item Leading-muon $\pt > 100 \gev$, and
\item Track multiplicity $\ntrk \ge 30$.
\end{itemize}

The validation regions are defined by inverting one or both of the above requirements. Explicitly,
the validation regions are split into two types:
\begin{itemize}
\item Leading muon satisfies $40 < \pt < 100 \gev$ without any requirement on $\ntrk$.
\item Leading-muon $\pt > 100 \gev$ and $\ntrk < 24$.
\end{itemize}
The validation regions are further split into bins in track multiplicity to test the background
estimation techniques described in the next section.

\section{Background Estimation}
\label{sec:backgrounds}
The backgrounds from Standard Model processes are divided into two categories for ease of estimation: processes where
the two muons come from correlated decay chains and processes that produce like-sign dimuons in uncorrelated decay chains.
Examples of correlated decay chains are the decays of $\ttbar$ events, where there is a fixed branching ratio
to obtain like-sign dimuons. The most likely scenario in \ttbar\ events
is where the leading isolated muon arises from the decay of a $W$ boson from one of the top quarks, and the subleading
muon of the same charge comes either from the semileptonic decay of a $b$ quark from the other top quark, or from the
sequential decay $b \rightarrow cX \rightarrow \mu X'$ of a $b$-quark from the same top quark.

The uncorrelated background estimates arise predominantly from the $W+$jets process, where the $W$ boson decay gives rise to the
leading isolated muon and the other muon arises from an in-flight $\pi / K$ decay, or the semileptonic decay of a $B$ or $D$ hadron.
Processes such as $Z$+jets, and single top in the $s$- and $t$-channels also give rise to uncorrelated backgrounds when the
leading isolated muon arises from the vector boson decay, and one of the jets gives rise to the second muon. The second muon is referred
to as a ``fake'' muon in the subsequent discussion of the uncorrelated background estimate.

\subsection{Correlated Background Estimates}
\label{sec:mcbkg}
The following sources of correlated backgrounds are considered: $\ttbar$ production, diboson production, and
single-top production in the $Wt$ channel. 
Each of the three correlated backgrounds is estimated from dedicated MC samples. The background from the $Wt$ process is small and it has been
merged with the $\ttbar$ background in the subsequent discussion and presentation.
Other possible sources such as $\ttbar W$ or $tZ$ production and backgrounds from charge misidentification of
muons were found to be negligible. 

The sources of uncertainty on the $\ttbar$ background are the choice of MC event generator and parton showering model, the amount of initial- and
final-state radiation (ISR/FSR), and the theoretical uncertainty on the production cross section. 
The $\ttbar$ cross section used is $\sigma_{\ttbar}= 238^{+22}_{-24}$~pb for a top-quark mass of 172.5~\gev. 
It has been calculated at approximate next-to-next-to-leading order (NNLO) in QCD with {\sc hathor} v1.2~\cite{hathor} using the MSTW 2008 90\% NNLO PDF sets.
It incorporates PDF+$\alpha_S$ uncertainties, according to the MSTW prescription~\cite{Martin:2009bu}, added in quadrature 
with the scale uncertainty and has been cross-checked with the calculation of Cacciari et al.~\cite{Cacciari} as implemented 
in {\sc top}\verb++++ v1.0~\cite{toppp}. 
The uncertainty on the parton showering model is assessed by comparing the nominal $\ttbar$ prediction with
a prediction made using a \powheg+\herwig\ sample. The generator uncertainty is assessed by comparing the \powheg\
prediction with predictions made using the \alpgen\ and \mcatnlo\ samples.
The ISR/FSR uncertainty is determined
by using the \acer~\cite{acermc} generator interfaced to \pythia, and by varying the ISR and FSR scale $\Lambda_{\mathrm{QCD}}$, 
as well as the ISR and FSR cutoff scale.
The effect of the top quark mass is studied by generating dedicated samples with top-quark masses of 170~\gev\ and 175~\gev\ and
is found to be negligible.

The diboson backgrounds have an uncertainty of 6\% on the production cross section~\cite{xsecdb} and a combined generator and parton-showering 
uncertainty of 24\%  based on comparisons between \sherpa\ and \powheg, and from renormalization and factorization scale variations~\cite{Aad:2012xsa}.

In addition to the uncertainties described above, uncertainties from the measurement of trigger efficiency, the
muon reconstruction and identification (including uncertainties due to muon \pt\ resolution), and the tracking efficiency are considered for each background along
with the uncertainty on the integrated luminosity (2.8\%). The total systematic uncertainties on the final background estimates from the 
different sources are summarized in Table~\ref{tab:systsummary}.

\begin{table}[tbp]
  \begin{center}
  \caption[Systematics]{ The systematic uncertainties on the event yields in the signal region for the different backgrounds and sources, in percent. 
The uncertainties on signal acceptance are also summarized in the table. }
    \begin{tabular}{ccccc}
      \hline \hline
      Source&              $\mu$+fake& \ttbar& Diboson& Signal \\ \hline
      Fake rate measurement&         34& & & \\
      Photon trigger&      13& & & \\
      Prompt correction&   18& & & \\
      ISR/FSR&              & 0.7&      & \\ 
      Parton showering&     & 9  &      & \\
      Generator&            & 11 &    24 & \\
      Cross section&        & 10 &    6 & \\ 
      Muon trigger&         & 1.2&   1.3& 1.3 \\
      Muon reconstruction&  & 2.0&   1.2& 2.3 \\
      Luminosity&           & 2.8&   2.8& 2.8 \\
      Tracking efficiency&   & 10&   10 & 10 \\
      Fiducial efficiency&  &    &       & 15 \\
      PDF (Acceptance)&     &    &       & 5\\ \hline
      Total&                41& 21&  27& 19\\ \hline \hline
    \end{tabular}
  \label{tab:systsummary}
  \end{center}
\end{table}

\subsection{Uncorrelated Background Estimates}
\label{sec:dtbkg}

The uncorrelated background is estimated from data by first measuring the probability for a track to be 
reconstructed as a muon in a control sample. This probability is then applied to data events with one muon
and at least one track to predict the number of dimuon events. This probability is referred to as a fake rate in the
subsequent discussion, and the background estimate is referred to as the $\mu+$fake background. 

The fake rate is measured in a control sample consisting of photon+jet events. These events are collected by a single-photon
trigger with a threshold at 40~\gev\ on the photon transverse momentum. The trigger is prescaled, and the collected dataset
corresponds to an integrated luminosity of 56~\ipb. The photon is required to have $\pt > 45 \gev$, and to satisfy the requirements of
Ref.~\cite{atlasphoton}.
The photon is also required to satisfy $E_{\mathrm{T}}^{\DeltaR \le 0.4} < 5 \gev$, where
$E_{\mathrm{T}}^{\DeltaR \le 0.4}$ is the sum of transverse energies of cells in the electromagnetic and hadronic calorimeters in a cone of 0.4 around the
photon axis (excluding the cells associated with the photon).
The denominator for the fake rate measurement is the number of events with one photon and at least one track. The track must satisfy all the requirements
imposed on an ID track associated with a muon as described in Sec.~\ref{sec:eventselection}. 
The track is required to be separated from the photon by ${\Delta R} > 0.4$.
The numerator is the subset of these events that have at least one muon passing all the criteria associated with the subleading muon as
described in Sec.~\ref{sec:eventselection}. In events with more than one track (muon), the track (muon) with the highest \pt\ is chosen.
The fake rate can have contributions from processes such as $W(\mu\nu)\gamma$ and $Z(\mu\mu)\gamma$ that produce prompt muons and 
bias the fake rate measurement. 
This prompt-muon bias is corrected by subtracting these contributions based on MC samples generated using \sherpa. The prompt muon correction ranges from approximately
1\% at muon $\pt = 15 \gev$ to 30\% at $\pt=100\gev$.

\begin{table}
  \begin{center}
  \caption[Validation]{ The predicted backgrounds in the validation regions compared to the number of observed data events.  The uncertainties shown on the total background are the combined statistical and systematic uncertainties. }
    \begin{tabular}{lccccc}
      \hline \hline
      \ntrk & \ttbar& Diboson& $\mu$+fake& Total& Data\\
      \hline
      \multicolumn{6}{c}{$40\gev<$ Leading-muon $\pt<100\gev$} \\
      \hline
      $\ntrk < 10$&         10000& 800& 20000& 31000 $\pm$ 4000& 28988 \\
      $10 \le \ntrk < 20$&    800&   3& 400&   1200  $\pm$ 100&   1103 \\
      $\ntrk \ge 20$&          16&   0.1& 6.8&  23    $\pm$ 3&    12 \\
      \hline
      \multicolumn{6}{c}{Leading-muon $\pt \ge 100\gev$} \\
      \hline
      $\ntrk<10$&            2400&  140& 2300&  4800 $\pm$ 600& 4428 \\
      $10\le \ntrk \le 11$&   190&    3&   76&     270  $\pm$ 31&  271 \\
      $12\le \ntrk \le 14$&   133&    1.1& 42&     176  $\pm$ 21&  167 \\
      $15\le \ntrk \le 19$&   60&     0.3& 17&      77   $\pm$ 9&   68 \\
      $20\le \ntrk \le 24$&   10&     0.1& 2.9&     13   $\pm$ 2&   13 \\
\hline \hline
    \end{tabular}
  \label{tab:validation}
  \end{center}
\end{table}

The fake rate is parameterized as a function of the \pt\ and $\eta$ of the track, and as a function of \ntrk.
Since the signal black hole models produce isolated photon events as well, the fake rate is measured by requiring $\ntrk <10$ to reduce
any potential signal contamination of this control sample. 
The \ntrk\ dependence is parameterized with a linear fit, and extrapolated for all events with $\ntrk > 10$. 
The average fake rate based on the criteria defined here is approximately 1\%. The fake rate is consistent with that obtained from photon+jet or $W$+jet MC samples.
The final $\mu+$fake background estimate
is obtained by selecting events with one muon satisfying the requirements of a leading muon, and one track of the same charge satisfying the
requirements associated with an ID track. These $\mu^\pm$track$^\pm$ events are then assigned a weight based on the fake rate calculated for the
track, and are then taken through the rest of the analysis chain in the same way as $\mu^\pm\mu^\pm$ events, with the track acting as the proxy for
the subleading muon. There is a correction to the fake estimate from overcounting due to $\ttbar$ and diboson events populating the $\mu^\pm$track$^\pm$ events
in data. This correction is estimated from MC simulation to be 2\% and is negligible compared to the systematic uncertainty on the fake estimate.

The uncertainties in the $\mu+$fake background estimate arise from the statistical uncertainties in measuring the fake rate, the
choice of the photon trigger used for the control sample, and the prompt-muon bias correction. The statistical
uncertainty in the fake rate is propagated to the final background estimate, along with the uncertainty in the fit parameters used to
parameterize the \ntrk\ dependence. The fake rate is remeasured using data collected by a single-photon trigger with a \pt\ threshold of 80~\gev, and the
background estimate is recalculated to assess the uncertainty due to the photon trigger. To assess the uncertainty due to the prompt correction,
the correction is varied by $\pm$15\% of its nominal value to obtain ``up'' and ``down'' fake rates. The final $\mu+$fake background estimate is calculated
with the up and down fake rates, and the larger variation from the nominal estimate is assigned as a systematic uncertainty. The choice
of $\pm$15\% is motivated by the uncertainties described in Ref.~\cite{atlaszgamma} that include experimental uncertainties on
photon reconstruction and identification, and theoretical uncertainties on the production cross sections of the $W\gamma/Z\gamma$ processes.
Table~\ref{tab:systsummary} shows the effect of the systematic uncertainties on the final $\mu+$fake background estimate.

\begin{figure}[tbp]
  \centering
   \includegraphics[width=\columnwidth]{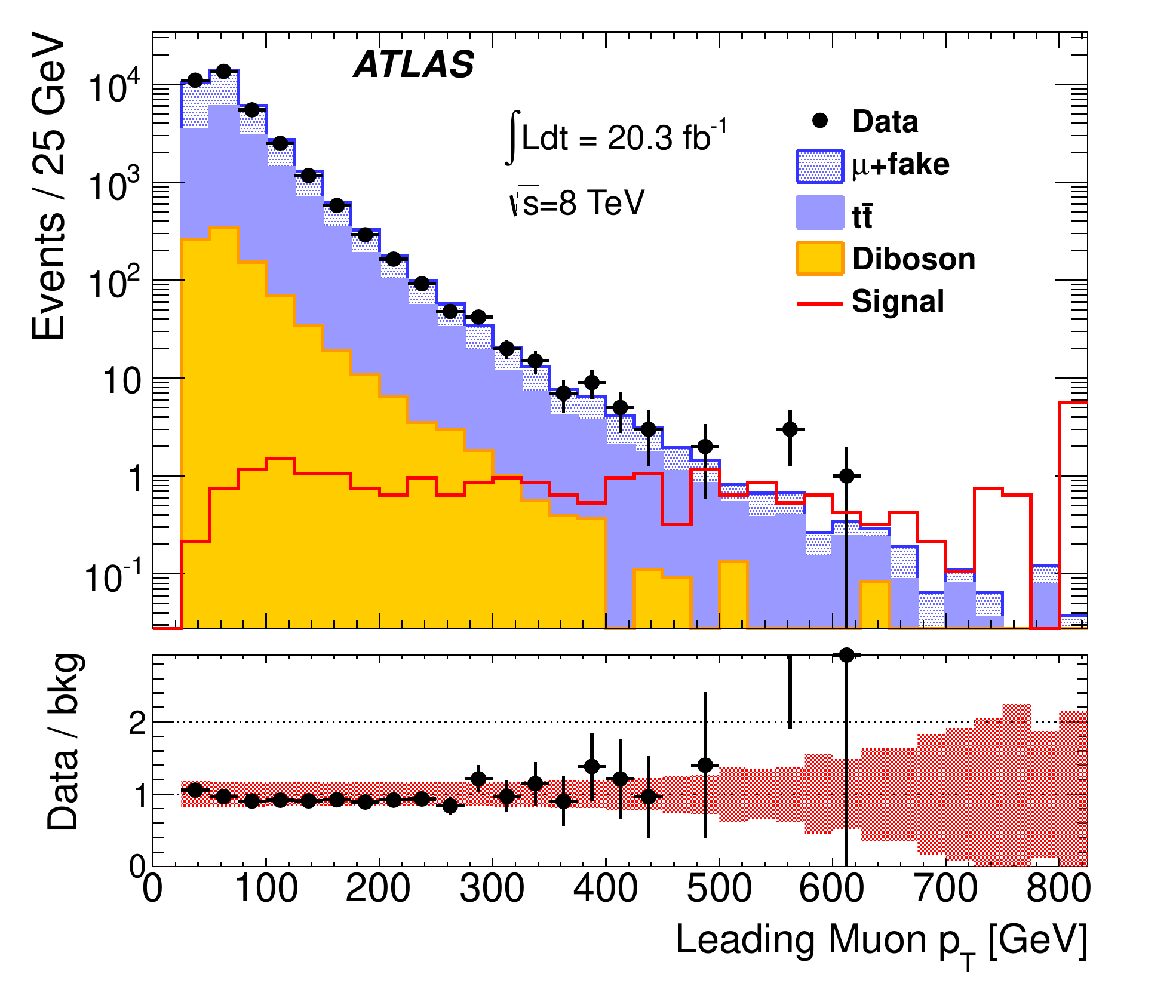}
  \caption{The leading-muon \pt\ distribution for the predicted background and observed data 
    for all like-sign dimuon events passing the preselection criteria. The background histograms are stacked. The signal
histogram is overlaid. The last bin along the $x$-axis shows the overflows.
The bottom panel shows the ratio of data to the expected background (points) and the total uncertainty
on the background (shaded area).}
  \label{fig:pt1}
\end{figure}

\begin{figure}[tbp]
  \centering
   \includegraphics[width=\columnwidth]{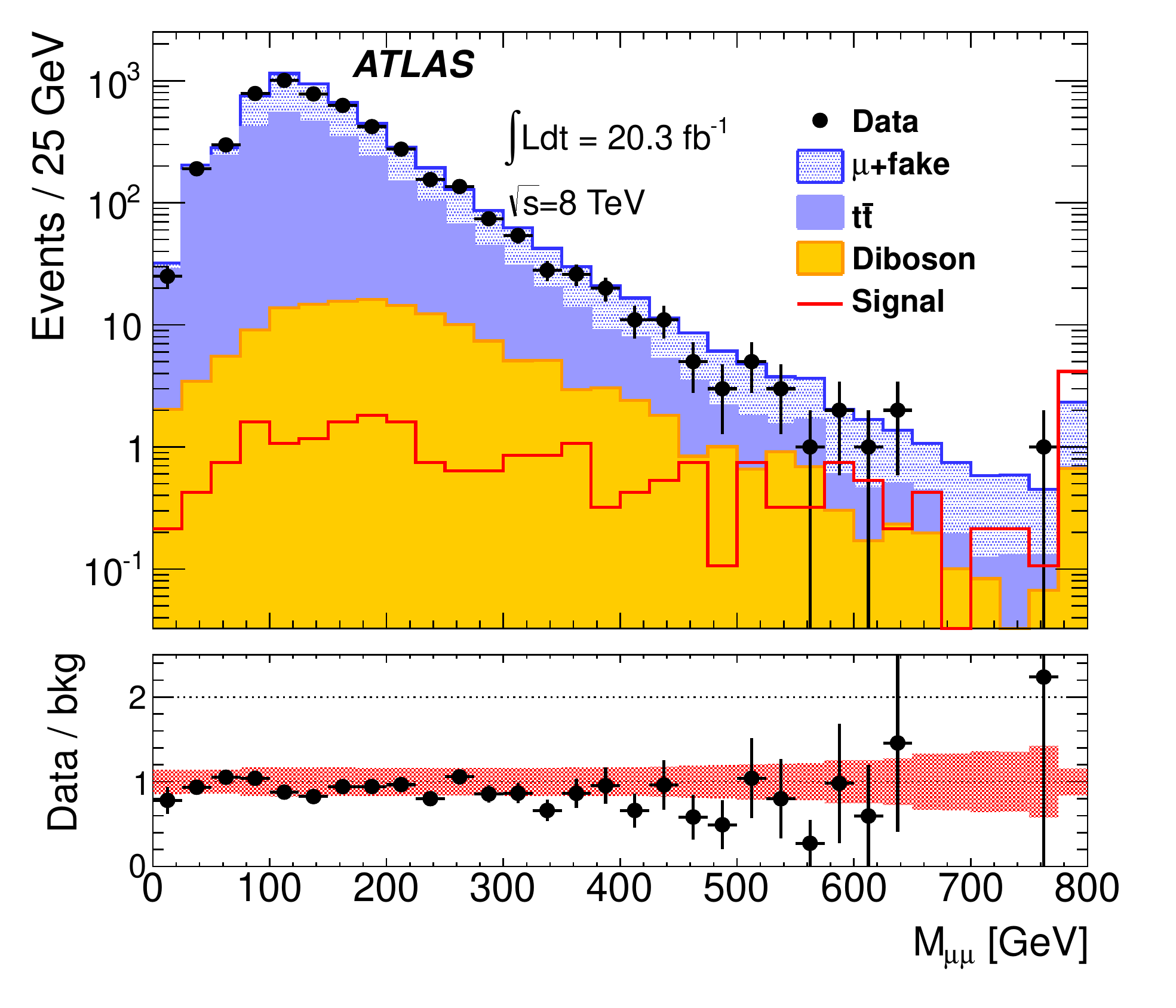}
  \caption{The dimuon invariant mass (M$_{\mu\mu}$) distribution for the predicted background and 
observed data for like-sign dimuon events where
the leading muon satisfies $\pt > 100 \gev$. The background histograms are stacked. The signal
histogram is overlaid. The last bin along $x$-axis shows the overflows. 
The bottom panel shows the ratio of data to the expected background (points) and the total uncertainty
on the background (shaded area). }
  \label{fig:cr100_mmm}
\end{figure}

\section{Results and Interpretation}
\label{sec:results}

The background estimation techniques described in the previous section are tested in the validation regions defined in
Sec.~\ref{sec:eventselection}. Table~\ref{tab:validation} shows the predicted backgrounds in the various validation regions and the observed yields. The signal contamination is negligible in all the validation regions. Overall, good agreement is observed between the prediction and the observation.

Figure~\ref{fig:pt1} shows the leading-muon \pt\ distribution for all like-sign dimuon events (satisfying the preselection). 
Figure~\ref{fig:cr100_mmm} shows the dimuon invariant mass distribution after imposing the $\pt > 100 \gev$ requirement on the leading muon.
Figure~\ref{fig:cr_ngt10_dphi} shows the distribution of the dimuon azimuthal separation, $\Delta\phi_{\mu\mu}$ 
for events with $\ntrk \ge 10$ and where the leading muon has $\pt > 100 \gev$.
Figure~\ref{fig:ntrk} shows the track multiplicity distribution, which shows good agreement between predicted backgrounds and data.
The predicted background and the observed data events in the signal region are shown in Table~\ref{tab:result}. The figures and the table show the
expected signal contribution from rotating black holes in a model with $n=4$, $\MTH = 5 \TeV$ and $\MD = 1.5 \TeV$.

\begin{figure}
  \centering
   \includegraphics[width=\columnwidth]{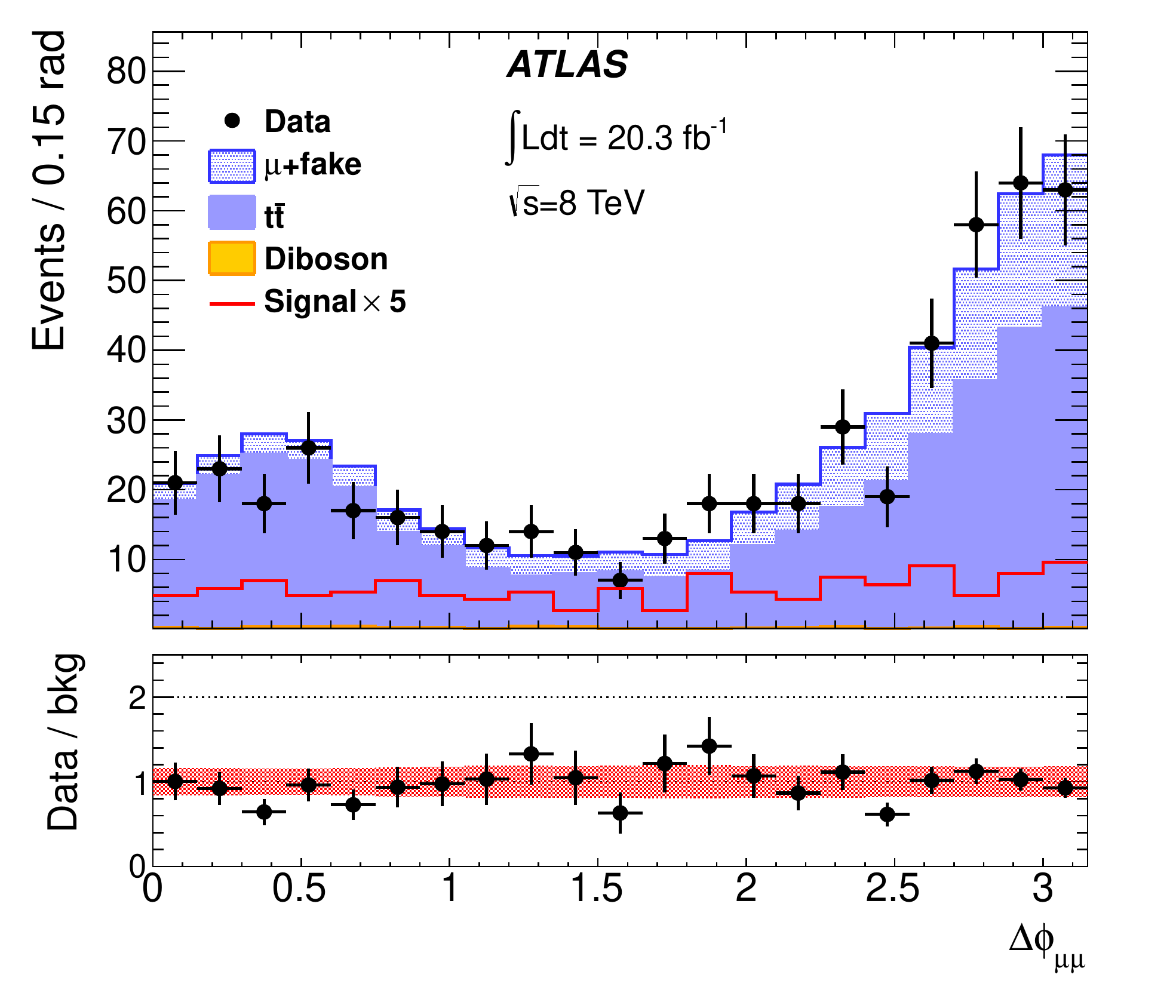}
  \caption{The dimuon azimuthal separation ($\Delta\phi_{\mu\mu}$) distribution for the predicted background and observed data 
for like-sign dimuon events where
the leading muon satifies $\pt > 100 \gev$, and with $\ntrk \ge 10$. The background histograms are stacked. The signal histogram is overlaid.
The bottom panel shows the ratio of data to the expected background (points) and the total uncertainty
on the background (shaded area). }
  \label{fig:cr_ngt10_dphi}
\end{figure}

\begin{figure}
  \centering
   \includegraphics[width=\columnwidth]{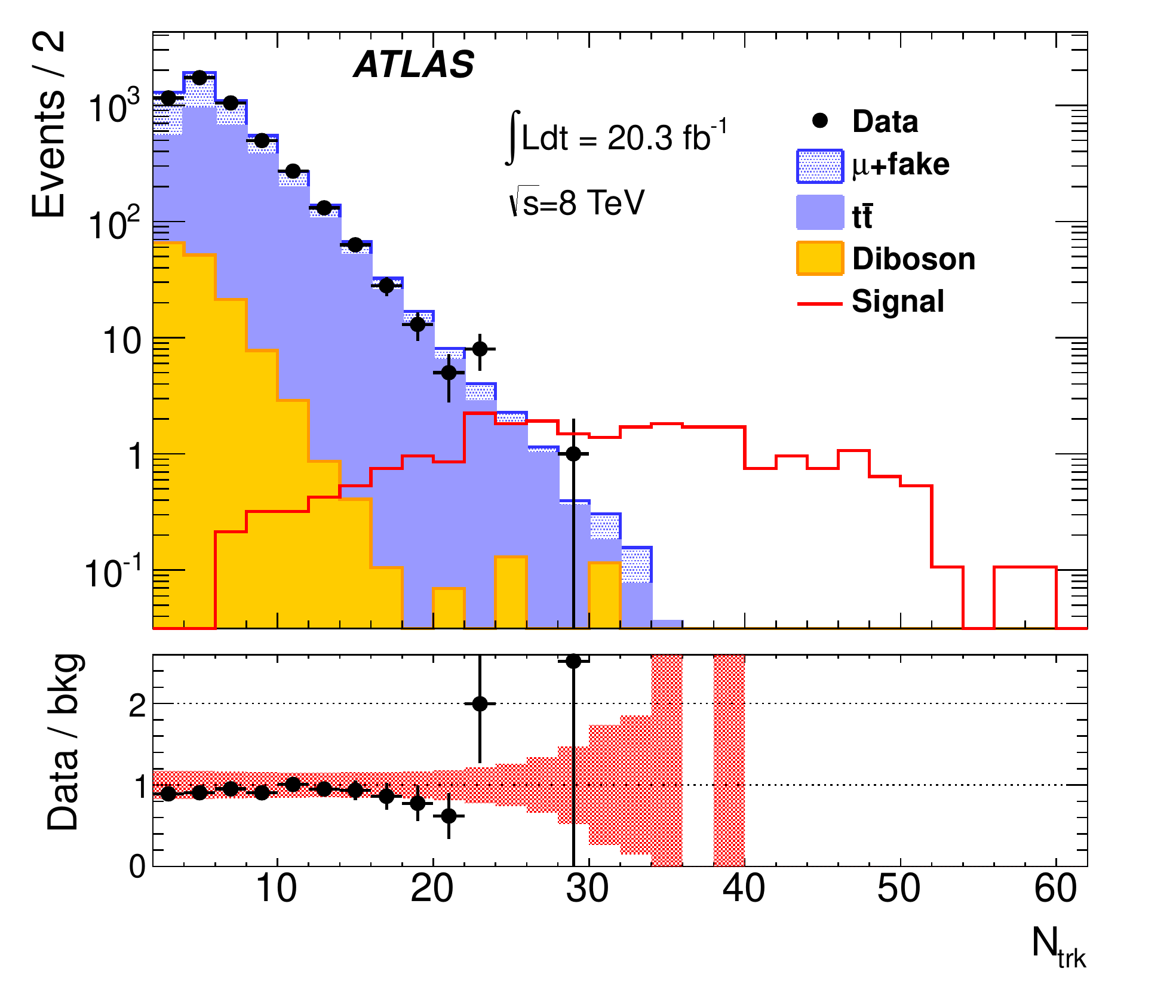}
  \caption{The track multiplicity (\ntrk) distribution ($\pt^{\mathrm{trk}} > 10 \gev$) for the predicted background and observed data 
    for events where the leading muon has $\pt > 100 \gev$. The background histograms are stacked. The signal
histogram is overlaid.
The bottom panel shows the ratio of data to the expected background (points) and the total uncertainty
on the background (shaded area).}
  \label{fig:ntrk}
\end{figure}

 \begin{table}
  \begin{center}
  \caption[Signal Region]{ The predicted background in the signal region compared to the number of observed events in data. The MC predictions are shown together with statistical and systematic uncertainties. The expected signal contribution from rotating black holes in a model with $n=4$, $\MTH = 5 \TeV$ and $\MD = 1.5 \TeV$ is also shown.}
    \begin{tabular}{ll}
      \hline \hline
      Source& Signal Region\\ \hline
      $\mu+$fake& 0.21 $\pm$ 0.09 $\pm$ 0.09\\
      \ttbar&  0.22 $\pm$ 0.08 $\pm$ 0.04\\
      Diboson& 0.12 $\pm$ 0.08 $\pm$ 0.03\\
      \hline
      Total& 0.55 $\pm$ 0.15 $\pm$ 0.10\\
      Data& 0\\
      \hline
      Signal& 14.2 $\pm$ 1.3 $\pm$ 2.7\\
      \hline
      \hline	
    \end{tabular}
  \label{tab:result}
  \end{center}
\end{table}


No events are observed in the signal region, which is consistent with the Standard Model prediction. This result is used to set 
upper limits on 
the number of events from non-Standard Model sources. The $CL_s$ method~\cite{cls} is used to calculate 95\% CL upper limits
on  $\sigma_{\mathrm{vis}} = \sigma \times BR \times A \times \epsilon$, where $\sigma_{\mathrm{vis}}$ is the visible cross section, 
$\sigma$ is the total cross section, $BR$ is the inclusive branching ratio to like-sign dimuons, $A$ is the acceptance, and $\epsilon$ is the reconstruction efficiency 
for non-Standard Model contributions in this final state in the signal region.  The observed 95\% CL limit on
$\sigma_{\mathrm{vis}}$ is 0.16~fb. The observed limit agrees well with the expected limit of 0.16~fb. 
The standard deviation ($\sigma$) bands on the expected limit at $1\sigma$ and $2\sigma$ are 0.15--0.22~fb and 0.15--0.29~fb, respectively.

Exclusion contours in the plane defined by \MTH\ and \MD\ for rotating and non-rotating black holes for $n$ = 2, 4, and 6 are obtained.
No theoretical uncertainty on the signal prediction is assessed, i.e. the exclusion limits are set for the exact benchmark 
models as described in Sec.~\ref{sec:samples}. 

The signal acceptance is measured from the event generator (truth) by imposing the following selections at the particle level.
Each event must have at least two true muons with $\pt > 15 \gev$ and $|\eta| < 2.4$, and the leading two muons in $\pt$ must have the same charge. 
The leading muon must satisfy
$\pt > 100 \gev$. The leading-muon truth-isolation ($I_{\mathrm{gen}}$) is defined as the sum of \pt\ of all charged particles with $\pt > 1 \gev$ within a cone of 
${\Delta R} = 0.2$ around the muon (excluding the muon). The leading muon is required to satisfy $I_{\mathrm{gen}} < 0.25\times \pt$. Each event must also
have at least 30 charged particles satisfying $\pt > 10 \gev$ and $|\eta|<2.5$. The ratio of events passing these selections at particle level to the total 
number of generated events gives the acceptance. The acceptance varies from 11\% to 0.2\% across the range of model parameters considered here.

The acceptance is then corrected to take into account detector effects. The correction factor, $\epsilon_{\mathrm{fid}}$, is defined as the ratio of number 
of events passing the selection criteria after full detector reconstruction to the number of events passing the acceptance criteria at the particle level.
The factor is found to be independent of the number of extra dimensions, and is
linearly dependent on $k =\MTH/\MD$. The linear dependence is assessed separately for rotating and non-rotating black holes by a fit to the
efficiency as a function of $k$. For rotating (non-rotating) signals  $\epsilon_{\mathrm{fid}}$ rises from 0.35 (0.3) for $k=1$ to 0.55 (0.65) for $k=3$.

\begin{figure}
  \centering
   \includegraphics[width=\columnwidth]{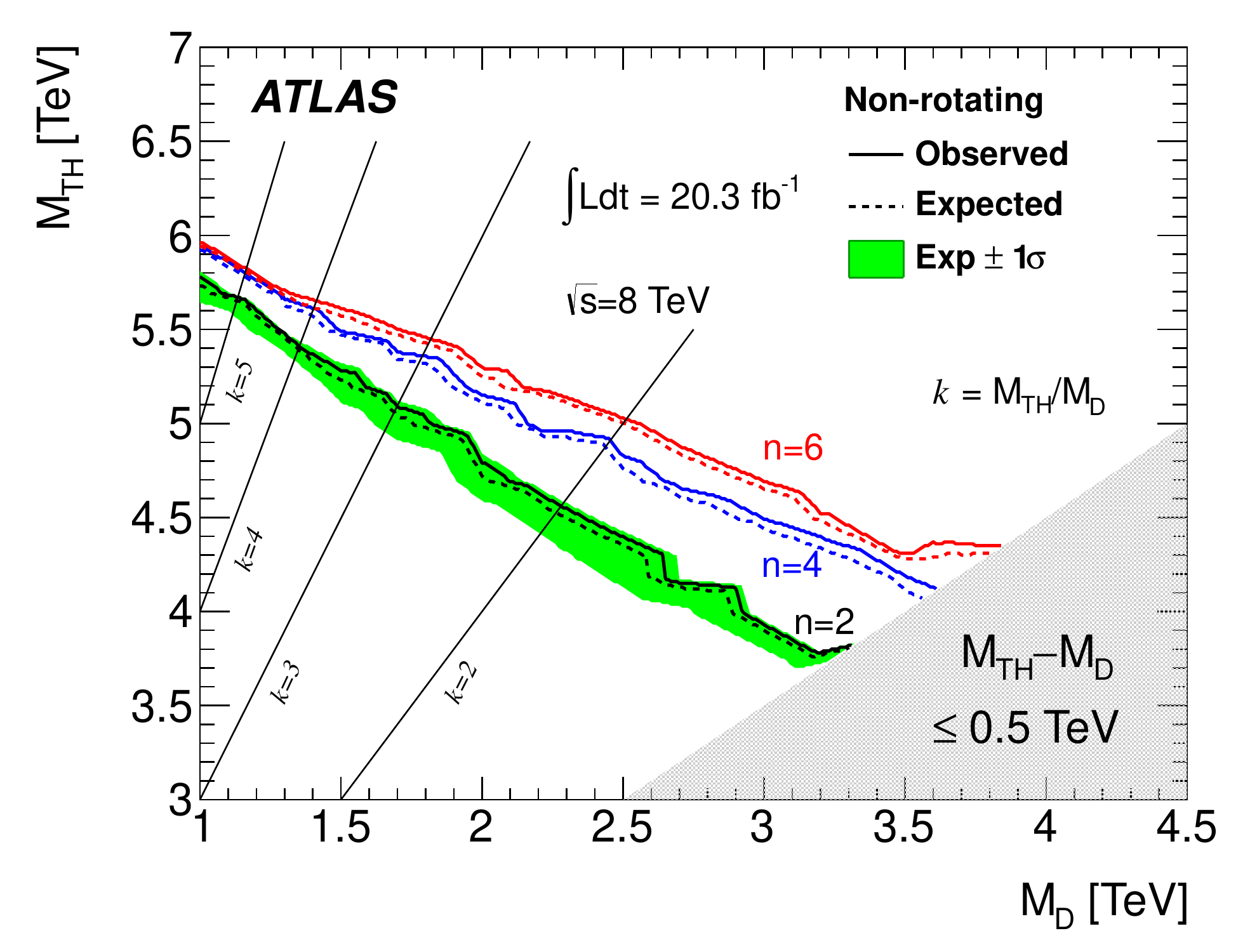}
  \caption{The 95\% CL exclusion contours for non-rotating black holes in models with $n$ = 2, 4, and 6.
The dashed lines show the expected exclusion contour,  the solid lines show the observed exclusion contour. The regions below the contours
are excluded by this analysis. The $1\sigma$ uncertainty on the expected limit for $n$ = 2 is shown as a band.
Lines of constant slope $k = \MTH/\MD$ = 2, 3, 4, and 5 are also shown. Only slopes $k\gg 1$ correspond to physical models.}
  \label{fig:bh1}
\end{figure}

The uncertainty on the signal prediction has the following components: the uncertainty on the $\epsilon_{\mathrm{fid}}$ fit parameters,
the uncertainty on luminosity, the uncertainty on acceptance due to the PDFs, the experimental uncertainty on acceptance due to muon trigger and
identification efficiencies, and the uncertainty due to tracking efficiency. 
The uncertainty on acceptance due to PDF was estimated by using the 40 error sets associated to the MSTW 2008 LO PDF set. In the signal region, at high \ntrk,
it is possible for small differences between the track reconstruction efficiencies in data and simulation to be magnified. The effect of any possible disagreement between
data and simulation is studied by artificially increasing the disagreement and probing the subsequent effect on the signal acceptance. A disagreement of 2\% in the
per-track reconstruction efficiency translates to a 5\% uncertainty in the signal acceptance for $\ntrk \ge 30$. 
As a conservative choice, a 10\% uncertainty on signal acceptance
is assigned to account for possible disagreements in data and simulation track reconstruction efficiency.
The uncertainties are summarized in Table~\ref{tab:systsummary}.

\begin{figure}
  \centering
   \includegraphics[width=\columnwidth]{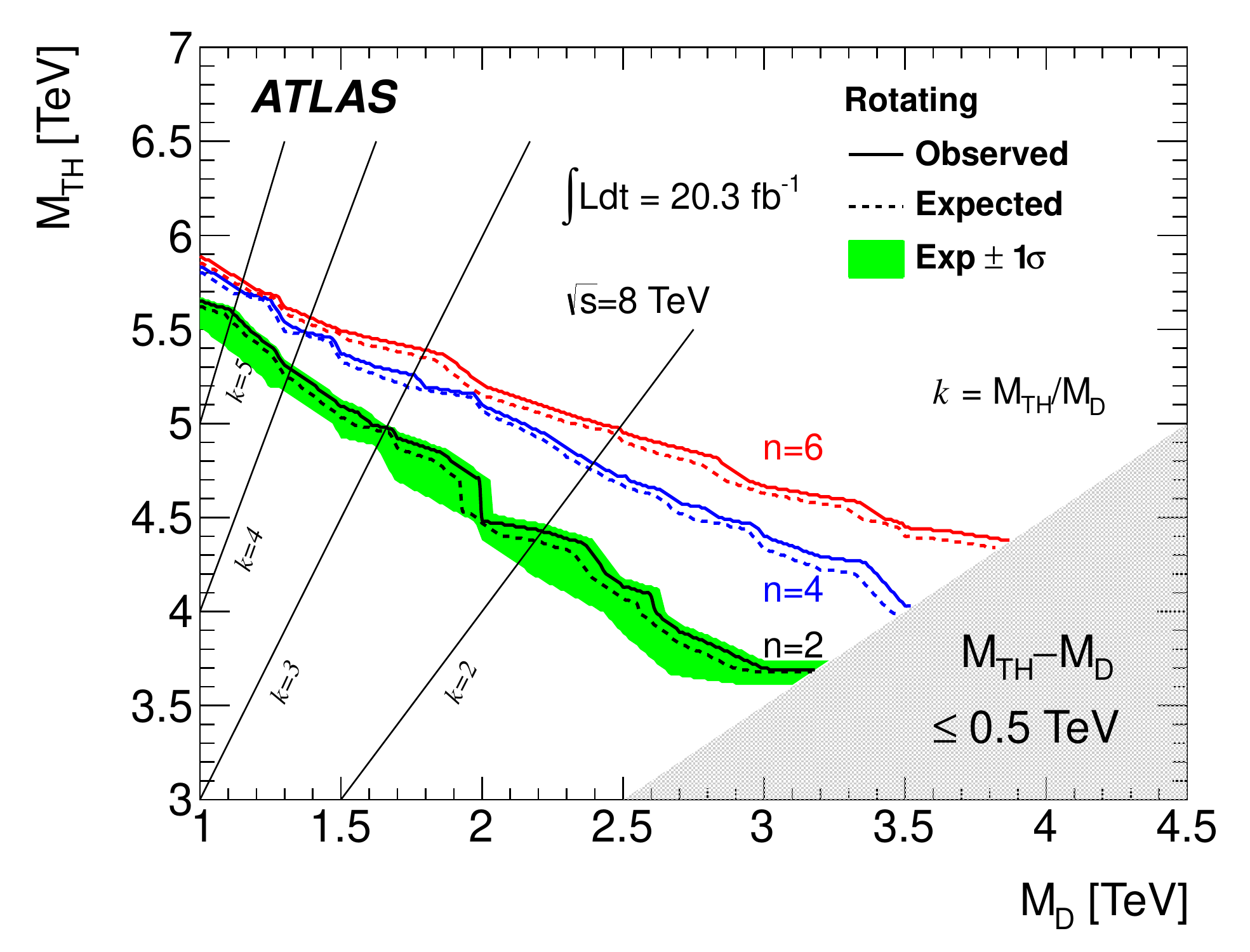}
  \caption{The 95\% CL exclusion contours for rotating black holes in models with $n$ = 2, 4, and 6.
The dashed lines show the expected exclusion contour, the solid lines show the observed exclusion contour. The regions below the contours
are excluded by this analysis. The $1\sigma$ uncertainty on the expected limit for $n$ = 2 is shown as a band.
Lines of constant slope $k = \MTH/\MD$ = 2, 3, 4, and 5 are also shown. Only slopes $k\gg 1$ correspond to physical models.}
  \label{fig:bh2}
\end{figure}

Figure~\ref{fig:bh1} shows the expected and observed exclusion contours for non-rotating
black holes for $n$ = 2, 4, and 6. Figure~\ref{fig:bh2} shows the same for rotating black holes. In both figures, 
the 1$\sigma$ uncertainty band on the expected limit is shown for $n$ = 2. For each value of $n$, the observed limit lies within
the 1$\sigma$ band.
Lines of constant slope ($k = \MTH/\MD$) of 2, 3, 4 and 5 are also shown. 
The semi-classical approximations used for black hole production and decay are expected to be valid only for large slopes. 
The effect of choosing a different set of PDFs for signal generation has been studied by considering the CT10 PDF set. 
The predicted cross sections with the CT10 PDF set are approximately 20\% higher, but this has a negligible impact on the exclusion 
contours due to the rapidly falling cross section with mass for black hole production.

The theory of large extra dimensions can be embedded into weakly-coupled string theory~\cite{Dimopoulos:2001qe,Gingrich:2008di}, giving rise to string balls
whose decay would be experimentally similar to the decay of black holes. Models of string balls have two additional parameters \MS\ and \gs, the string
scale and the string coupling constant respectively, in addition to \MTH, \MD, and $n$. \blackmax\ is used to simulate the production and decay of string
balls, and to obtain exclusion contours in the plane defined by \MTH\ and \MS. Following Ref.~\cite{Gingrich:2008di}, the values of \gs\ and \MD\ are set by
$\gs^{2} = 1/5^{\frac{n+2}{n+1}}$, and $\MD= 5^{\frac{1}{n+1}} \MS$.
The exclusion contour in the \MTH--\MS\ plane for string balls is shown in Fig.~\ref{fig:sb} for models 
with $n$ = 6 (where $\gs=0.40$ and $\MD = 1.26\MS$). 
Table~\ref{tab:summary} shows the summary of lower limits placed on the mass of microscopic black holes 
and string balls for $\MD=1.5 \TeV$ for different values of $n$.

\begin{figure}
  \centering
   \includegraphics[width=\columnwidth]{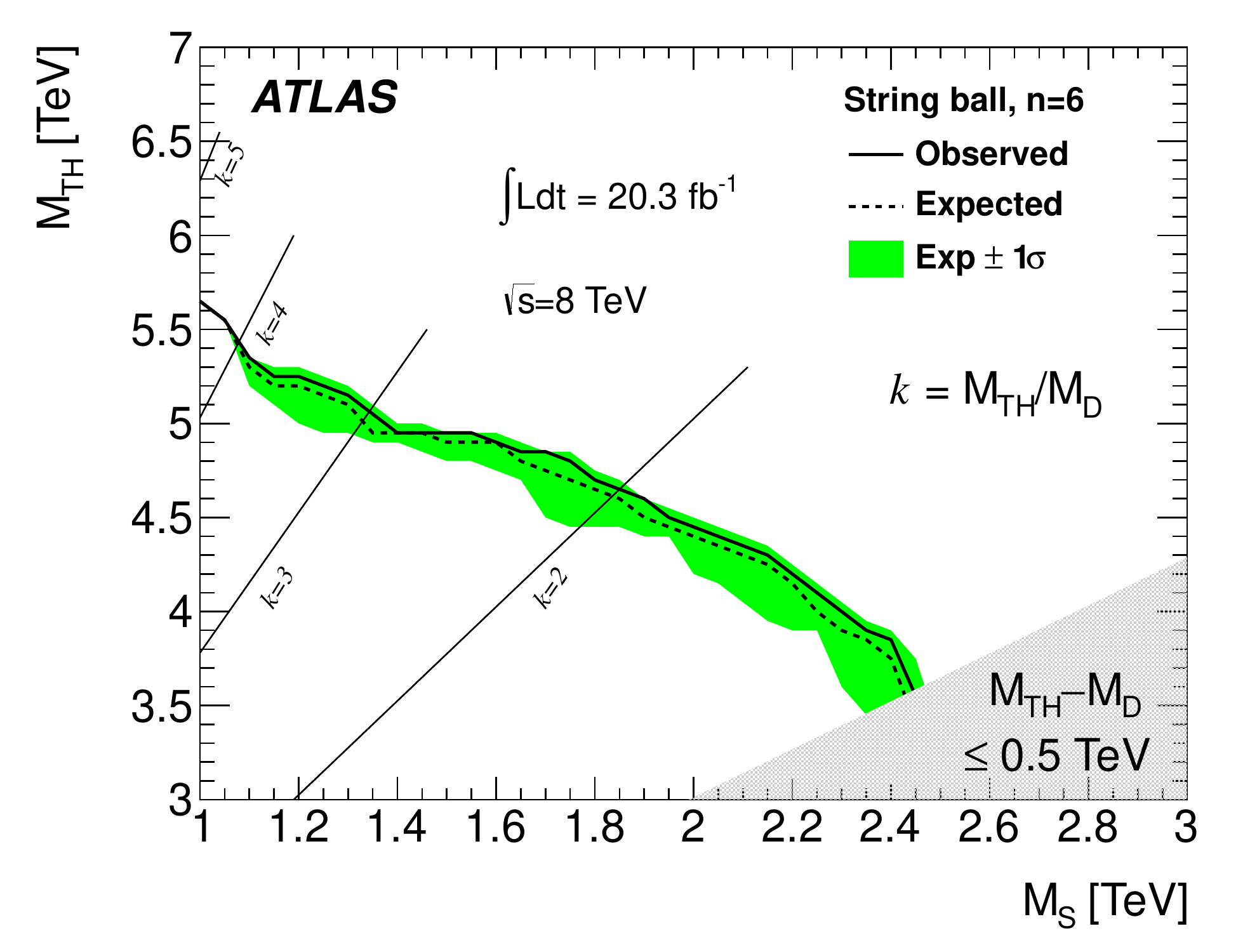}
  \caption{The 95\% CL exclusion contours for string balls in models with $n$ = 6.
The dashed line shows the expected exclusion contour with the
$1\sigma$ uncertainty shown as a band.  The solid line shows the observed exclusion contour. The region below the contour
is excluded by this analysis. Lines of constant $k = \MTH/\MD$ = 2, 3, 4, and 5 are also shown.  Only the values $k\gg 1$ correspond to physical models.}
  \label{fig:sb}
\end{figure}

\section{Conclusions}
\label{sec:conclusions}

\begin{table}
  \begin{center}
\caption[Exclusion Summary]{The lower limits on $\MTH$ at 95\% CL are summarized for different models. $\MD$ is fixed at 1.5~\TeV. For the string ball model, 
$\gs=0.40$ and $\MS = \MD/1.26 = 1.2~\TeV$.}
  \begin{tabular}{lcccc}
  \hline \hline
  Model&     &     $n$& &  $\MTH$[\TeV]$\ge$ \\ \hline
  Non-rotating black hole& & 2& &  5.3\\
  Non-rotating black hole& &  4& &  5.6\\
  Non-rotating black hole& &  6& &  5.7\\ \hline
  Rotating black hole& &  2& &  5.1\\
  Rotating black hole& &  4& &  5.4\\
  Rotating black hole& &  6& &  5.5\\ \hline
  String ball& &  6& &  5.3\\
 \hline
 \hline
\end{tabular}
\label{tab:summary}
\end{center}
\end{table}

A search for microscopic black holes has been carried out using 20.3~\ifb\ of data collected by the ATLAS detector in 8~\TeV\
proton--proton collisions at the LHC. No excess of events over the Standard Model background expectations is observed  in the final state with a like-sign dimuon pair
and high track multiplicity. Exclusion contours in the plane of the fundamental Planck scale \MD\ and the threshold mass
\MTH\ of black holes are shown and a limit of 0.16~fb at 95\% CL is set on the visible cross section for any new physics in the signal region defined by a like-sign dimuon pair and high track multiplicity selection.

\acknowledgments

We thank CERN for the very successful operation of the LHC, as well as the
support staff from our institutions without whom ATLAS could not be
operated efficiently.

We acknowledge the support of ANPCyT, Argentina; YerPhI, Armenia; ARC,
Australia; BMWF and FWF, Austria; ANAS, Azerbaijan; SSTC, Belarus; CNPq and FAPESP,
Brazil; NSERC, NRC and CFI, Canada; CERN; CONICYT, Chile; CAS, MOST and NSFC,
China; COLCIENCIAS, Colombia; MSMT CR, MPO CR and VSC CR, Czech Republic;
DNRF, DNSRC and Lundbeck Foundation, Denmark; EPLANET, ERC and NSRF, European Union;
IN2P3-CNRS, CEA-DSM/IRFU, France; GNSF, Georgia; BMBF, DFG, HGF, MPG and AvH
Foundation, Germany; GSRT and NSRF, Greece; ISF, MINERVA, GIF, DIP and Benoziyo Center,
Israel; INFN, Italy; MEXT and JSPS, Japan; CNRST, Morocco; FOM and NWO,
Netherlands; BRF and RCN, Norway; MNiSW, Poland; GRICES and FCT, Portugal; MERYS
(MECTS), Romania; MES of Russia and ROSATOM, Russian Federation; JINR; MSTD,
Serbia; MSSR, Slovakia; ARRS and MIZ\v{S}, Slovenia; DST/NRF, South Africa;
MICINN, Spain; SRC and Wallenberg Foundation, Sweden; SER, SNSF and Cantons of
Bern and Geneva, Switzerland; NSC, Taiwan; TAEK, Turkey; STFC, the Royal
Society and Leverhulme Trust, United Kingdom; DOE and NSF, United States of
America.

The crucial computing support from all WLCG partners is acknowledged
gratefully, in particular from CERN and the ATLAS Tier-1 facilities at
TRIUMF (Canada), NDGF (Denmark, Norway, Sweden), CC-IN2P3 (France),
KIT/GridKA (Germany), INFN-CNAF (Italy), NL-T1 (Netherlands), PIC (Spain),
ASGC (Taiwan), RAL (UK) and BNL (USA) and in the Tier-2 facilities
worldwide.

\bibliography{paper}

\appendix 

\clearpage
\onecolumngrid
\clearpage
\begin{flushleft}
{\Large The ATLAS Collaboration}

\bigskip

G.~Aad$^{\rm 48}$,
T.~Abajyan$^{\rm 21}$,
B.~Abbott$^{\rm 112}$,
J.~Abdallah$^{\rm 12}$,
S.~Abdel~Khalek$^{\rm 116}$,
O.~Abdinov$^{\rm 11}$,
R.~Aben$^{\rm 106}$,
B.~Abi$^{\rm 113}$,
M.~Abolins$^{\rm 89}$,
O.S.~AbouZeid$^{\rm 159}$
% ...
\bigskip
\\
% ...
$^{11}$ Institute of Physics, Azerbaijan Academy of Sciences, Baku, Azerbaijan\\
$^{12}$ Institut de F{\'\i}sica d'Altes Energies and Departament de F{\'\i}sica de la Universitat Aut{\`o}noma de Barcelona, Barcelona, Spain\\
% ...
$^{21}$ Physikalisches Institut, University of Bonn, Bonn, Germany\\
% ...
$^{48}$ Fakult{\"a}t f{\"u}r Mathematik und Physik, Albert-Ludwigs-Universit{\"a}t, Freiburg, Germany\\
% ...
$^{89}$ Department of Physics and Astronomy, Michigan State University, East Lansing MI, United States of America\\
% ...
$^{106}$ Nikhef National Institute for Subatomic Physics and University of Amsterdam, Amsterdam, Netherlands\\
% ...
$^{112}$ Homer L. Dodge Department of Physics and Astronomy, University of Oklahoma, Norman OK, United States of America\\
$^{113}$ Department of Physics, Oklahoma State University, Stillwater OK, United States of America\\
% ...
$^{116}$ LAL, Universit{\'e} Paris-Sud and CNRS/IN2P3, Orsay, France\\
% ...
$^{159}$ Department of Physics, University of Toronto, Toronto ON, Canada\\
% ...
\end{flushleft}


\end{document}